\documentclass{pasj00}
\usepackage{graphicx}
\usepackage{amsthm}

\begin{document}

\author{Fabiola M. A. Ribeiro \and Marcos P. Diaz}
\affil{Instituto de Astronomia, Geof\'{i}sica e Ci\^{e}ncias Atmosf\'{e}ricas,
Universidade de S\~{a}o Paulo,\\05508-900, S\~{a}o Paulo, SP, Brazil}
\email{fabiola@astro.iag.usp.br}
\title{Tomographic Simulations of Accretion Disks in Cataclysmic Variables - Flickering and Wind}

\maketitle

\begin{abstract}
Both continuum and emission line flickering are phenomena directly associated with the mass accretion process. In this work we simulate accretion disk Doppler maps including the effects of winds and flickering flares. Synthetic flickering Doppler maps are calculated and the effect of the flickering parameters on the maps is explored.
Jets and winds occur in many astrophysical objects where accretion disks are present. Jets are generally absent among the cataclysmic variables (CVs), but there is evidence of mass loss by wind in many objects. CVs are ideal objects to study accretion disks and consequently to study the wind associated with these disks.
We also present simulations of accretion disks including the presence of a wind with orbital phase resolution. Synthetic H$\alpha$ line profiles in the optical region are obtained and their corresponding Doppler maps are calculated. The effect of the wind simulation parameters on the wind line profiles is also explored. From this study we verified that optically thick lines and/or emission by diffuse material into the primary Roche lobe are necessary to generate single peaked line profiles, often seen in CVs. The future accounting of these effects is suggested for interpreting Doppler tomography reconstructions.
\end{abstract}

%\keywords{accretion, accretion disks --- binaries: close --- cataclysmic variables}
\KeyWords: accretion, accretion disks --- stars: binaries: close --- stars: cataclysmic variables

\section{Introduction}

Cataclysmic variables (CVs) are close binary systems composed by a white dwarf (primary) and a red dwarf or subgiant star (secondary). The secondary star fills its Roche lobe and matter is transferred to the white dwarf. Due to the angular momentum of the system (and if the magnetic field of the primary is not strong enough) there is the formation of an accretion disk around the primary.

The flickering (rapid variability) is observed as stochastic fluctuations in the emitted radiation, with timescales ranging from seconds to tenth of minutes and amplitudes from cents of magnitude to more than one magnitude. Flickering is not exclusive to CVs, but it is also observed in symbiotic stars (e.g. CH Cyg, \cite{Mik90}), X-ray binaries (\cite{Rey07}, \cite{Mal03}, \cite{Bap02}) and pre-main sequence stars (\cite{Ken00}, \cite{Cla05}). The flickering cannot be related only to accretion disks as it is observed in magnetic systems \citep{Tha50}; an early model to describe the flickering in the polar AM Her was proposed by \citet{Pan80}. On a wide perspective, the presence of flickering seems associated to the process of mass accretion itself.

The first CV where flickering was observed is UX UMa \citep{Lin49}. Since then, many studies were made aiming to locate the flickering source region on many objects. The flickering can be originated in many regions of the system. \citet{War71} observed that the U Gem flickering disappeared during the eclipse, and as the inner parts of the disk were not occulted during the eclipse, the flickering source was associated with the hot spot. \citet{Vog81} observed that OY Car flickering persisted during the hot spot eclipse, the same behavior was observed on HT Cas by \citet{Pat81} and Z Cha by \citet{Woo86}. \citet{Hor85} identified the flickering in some eclipsing systems as originated in the inner part of the accretions disk, \citet{Hor94} also associated the OY Car flickering with this disk region. \citet{Bru96} associated the flickering source in Z Cha with the region near to the white dwarf, but with other flickering sources appearing in other photometric states. 

As the CVs cannot be spatially resolved, indirect imaging methods are used. One of these methods is the Doppler tomography \citep{Mar88}, where the system emissivity is reconstructed from the observed line profiles. \citet{Dia01} proposed an extension of the Doppler tomography, the flickering tomography, where the line profile variability is mapped. In this method, we start calculating the variance of the observed line profiles into phase bins. Then, the instrumental, orbital, and secular components are subtracted from such variance, remaining only the intrinsic variance component. The instrumental component of the variance is calculated as a combination of the Poisson noise plus a Gaussian readout noise. The orbital component contribution is considered negligible if the data is sampled into small phase bins. The secular component is the remaining long term variability of the system. The effect of this later component was found to be negligible by replacing the long term average profiles by profiles computed from individual runs. From these intrinsic variance line profiles Doppler tomograms are calculated, mapping the line emission variability in the system. This technique was applied to V442 Oph \citep{Dia01}, where an isolated flickering source was not identified, and V3885 Sgr \citep{Rib07}, where the main flickering source seems to be associated with reprocessed UV radiation on the illuminated face of the secondary star.

The association between accretion disks and jets is observed in many astrophysical objects, as young stellar objects (YSO), low-mass and high-mass X-ray binaries, microquasars and active galactic nuclei. Concerning disks around white dwarfs, jets are observed in super-soft binaries (SSS) and some V Sge systems. Among the CVs, jets are not detected, except during the evolution of classical novae explosions \citep{Ret04}, where the jet formation is related to the additional energy available due to the nuclear reactions on the primary surface. \citet{Sok04} explained the absence of jets in CVs using the jet launching thermal model \citep{Tor84}, originally developed to explain jets in YSO. According to these authors, the CVs accretion rates are not high enough to produce jets. On the other hand, some CVs have mass loss by wind, and the mechanism that best describes the wind production in these systems is the line opacity (line driven winds) \citep{Pro05}.

Ultraviolet observations allowed the identification of mass loss by wind in CVs. UV ressonant lines (e.g. CIV $\lambda$1548, $\lambda$1551, NV $\lambda$1239, $\lambda$1243, Si $\lambda$1397) are observed with a component in absorption displaced to the blue in classical novae remnants and nova-like systems with high accretion rate. The fact that wind is detected only in systems with high accretion rates relates the wind emission to the accretion process. The P-Cygni profiles are observed only in systems where the orbital inclination is approximately lower than 65\fdg (eg. \cite{War95, Shl93}), indicating that the geometry of the line formation region could be bipolar. Models from \citet{Dre87} indicate that, even if the emission of the accretion disk is bipolar, the wind also must be a bipolar geometry.

Most of the studies about winds in CVs are based on UV ressonant lines. They basically follow two research lines: cinematic models fitted to observed profiles \citep{Lon02}, or hydrodynamic numerical modeling \citep{Pro99}. A major disagreement between the modeled line profiles and the observed ones appears in the fit of the P-Cyg emission component. Most of the wind models make use the standard disk model (\cite{Sha73}, \cite{Pri81}). Disk models with temperature given by the standard model are optically thick in Balmer lines, but the outer parts could be optically thick in Balmer lines and optically thin in the continuum \citep{Wil80}. Following this model the lines must be formed in the outer parts of the disk, but the observations show that the disk line emissivity are centrally concentrated. Absorption lines are obtained from optically thick disk simulations under LTE while NLTE effects must be considered to simulate emission lines. \citet{Ko96} presented a model where the accretion disk vertical structure and NLTE effects are considered, where a chromosphere is superposed to a standard model disk. A selfconsistent model for accretion disk extended atmosphere is still an open issue.

The main wind features are detected in UV, but wind effects can also appear in the optical region. In a brief survey, \citet{Kaf04} noticed P-Cyg profiles in the HeI $\lambda$5876, HeI $\lambda$7065 and H$\alpha$ line profiles of BZ Cam, Q Cyg, HR Del, DI Lac, BT Mon and AC Cnc. They presented a few spectra of each object showing P-Cyg features and extended H$\alpha$ line wings, that are signatures of mass loss by wind. Studies analyzing the behavior of these features with the orbital phase in the optical region were not presented until now. Disk spectral synthesis and eclipse mapping studies suggest that gas in the primary Roche lobe (e.g. \cite{Bap01}) contribute with scattering free-free and recombination continuum photons.

The presence of wind in CVs is a hypothesis raised to explain the emission at low velocities seen in many Doppler maps and the single peaked line profiles observed even on systems were a double peak line profile is expected (e.g. high inclination systems) \citep{Mur97}.

\section{Accretion Disk Simulations}

The simulation procedure is divided into two parts. First, a steady accretion disk emission is simulated, later, the flickering and/or wind is added. The simulated disk is geometrically thin and limited by the white dwarf at the inner rim and by the tidal radius at the outer rim. The emissivity of the disk is assumed to follow a power law with radial dependence
\begin{equation}
 F_j=a {r_j}^b \left( \frac{r_j dr d\theta}{d\lambda} \right) \cos(i)
\end{equation}
where $F_j$ is each element flux, $r_j$ its radial coordinate, $\theta$ its angular coordinate, $a$ the power law proportionality constant, $b$ the power law index, $i$ the orbital inclination and $\lambda$ the emission wavelength calculated from the velocity Doppler shift calculated from equation 2.
For each disk element the emission is calculated, the corresponding velocity is computed (eq. 2) and the contribution of each disk element is put into the spectra.
\begin{displaymath}
v(r,\theta) = \gamma +
\end{displaymath}
\begin{equation}
\left( v_K + \frac{2\pi}{P}
\sqrt { \left( \frac{a_K}{1-q} - r \cos(\theta) \right) ^2 + \left( r \sin(\theta) \right) ^2 }  \right)
\end{equation}
\begin{displaymath}
\times \sin(\theta - 2 \pi\phi) \sin(i)
\end{displaymath}

In equation 2, $\gamma$ is the systemic velocity, $v_K$ the Keplerian disk velocity, and $a_K$ the system's stellar components separation.
This procedure is repeated until all elements of the disk were computed. There is also the possibility of simulate a hot spot or a boundary layer. These regions are geometrically restricted on the disk surface and their emissivity is given as a fraction of the total disk emissivity.

Aiming to test our simulation method, a disk with an enhanced emission in a region similar to the hot spot at the outer parts of the disk and a disk with an enhanced emission in the inner disk parts were simulated. From the synthetic spectra generated by the simulations, Doppler maps were constructed. In the first case, the hot spot-like emission appears on the expected region of the synthetic Doppler maps, but the emission of the inner parts of the disk does not appear on the tomograms. From this we conclude that the Doppler tomography method is appropriated to study emission from the outer parts of the accretion disks but not for emission from the inner parts of the disk. In this later case, the information concentrated on a small region in position space is spread on a large region on the velocity space, due to the Jacobian of this transformation.

\subsection{Flickering Simulations}

The emission line flickering is simulated as a set of discrete flares on the accretion disk. Each flare is generated at a random position inside a pre-defined region (the hot spot, for example), allowing us to simulate flickering from different regions of the disk  or from the secondary star. The flickering flare has a random intensity between a maximum intensity (simulation parameter) and zero. This maximum intensity is given in disk total line flux units, aiming to quantify the flare energy.

When flickering is observed in an object, the resulting data comprises a sum of all flickering flares events during each integration time. Aiming to reproduce this observational effect on the simulations, the flickering flares are represented along a temporal grid with bins much smaller than the integration time. The sum along many bins is made to synthetize the effect of flickering on each spectrum. As the flickering flares are not instantaneous events but they present a temporal evolution, the flares are simulated as having instantaneous rise and exponential decay
\begin{equation}
I(t)_{flare} = I(t_0)_{flare} e^{-t/\tau}
\end{equation}
where $I(t)_{flare}$ is the flare intensity on each bin, $I(t_0)_{flare}$ its intensity on the previous bin, $\tau$ the characteristic decay time and $t$ the temporal variable. The study of CVs light curves by \citet{Bru89} indicate an assimetry on the flickering flare temporal evolution, with the flare rise faster than its decay. A fiducial flare description is not a critical issue on our simulations as the integration times are often longer than the individual flare timescale.

As the flickering flares are not produced with a fixed periodicity, an occurrence probability is attributed to the flare on each time bin. This probability together with the integration time give us the averaged number of expected flares on each simulated line profile. The time of occurrence of the flare inside a particular temporal bin is a random value.

This procedure is repeated for all bins inside the integration time. The Doppler shift of the flares are calculated based on the position of each flare, assuming a Keplerian disk, and the flare contribution is added to the line profile. Aiming to simulate line profiles with noise, we introduced a Gaussian noise in the simulations, this noise is quantified by a continuum signal-to-noise ratio. The standard deviation of the noise distribution is calculated by the ratio between the continuum intensity and the signal-to-noise ratio desired for the line profile. The introduced noise is symmetric around zero average and it is added to each wavelength resolution element of the resulting line profile.

An averaged disk line profile and its corresponding RMS is shown in figure 1. Here the flickering simulations were held without noise. The disk was simulated on a 200x200 bins spatial grid. The addopted simulation parameters were the followings: orbital period of 5 hours, primary mass of 1 solar mass, mass ratio of 0.7 and integrated line flux of $9\times10^{-11}$ erg s$^{-1}$ cm$^{-2}$. The flickering source region is a small region restricted between 0.75 and 0.85 disk radii units, centered at an angle of 120$\fdg$ (measured from the line that join the stellar centres) and with an angular range of 1$\fdg$. This region has a location similar to the one of the impact of the stream on the disk considering free particles and a disk radius of approximately 2/3 of the primary Roche lobe radius. The flickering temporal evolution is calculated in a grid with 5 times more elements than the predicted number of flares on each integration time. The total decay time addopted for each flare was 1 minute. The maximum amplitude of the flares into each exposition was 10\% of the disk total flux. The flare occurrence mean frequency is 0.025 s$^{-1}$, which corresponds to a flare at each 40 seconds on average. The relative intensity between the line profile and its RMS is dependent on these parameters. The flickering was restricted to a hot-spot-like region, the effect of this assumption is seen on figure 1, as the RMS profile lacking high velocity wings. The noise on the RMS profile is produced by the finite number of phase bins where the variance is calculated. This behavior tends to be minimized when a large number of phase bins is used.

\begin{figure}
 \begin{center}
 \FigureFile(80mm,80mm){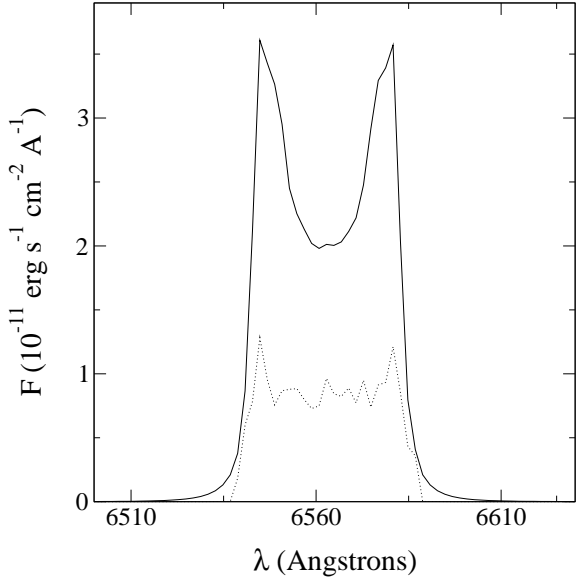}
 \end{center}
 \caption{Averaged simulated line profile (solid line) and its flickering RMS (dotted line) line profile. The simulations parameters are given on the text.}
 \label{fig1}
\end{figure}

Other simulations with flickering originating on the hot spot region were held. The simulation parameters were varied one by one to verify the effect on the synthetic tomograms. The addopted simulation parameters were the same of the last paragraph, but with an integrated flux of $10^{-9}$ erg s$^{-1}$ cm$^{-2}$, flickering maximum amplitude of 1\% of the total line flux and including noise in the line profiles.

\begin{figure}
 \begin{center}
 \FigureFile(60mm,200mm){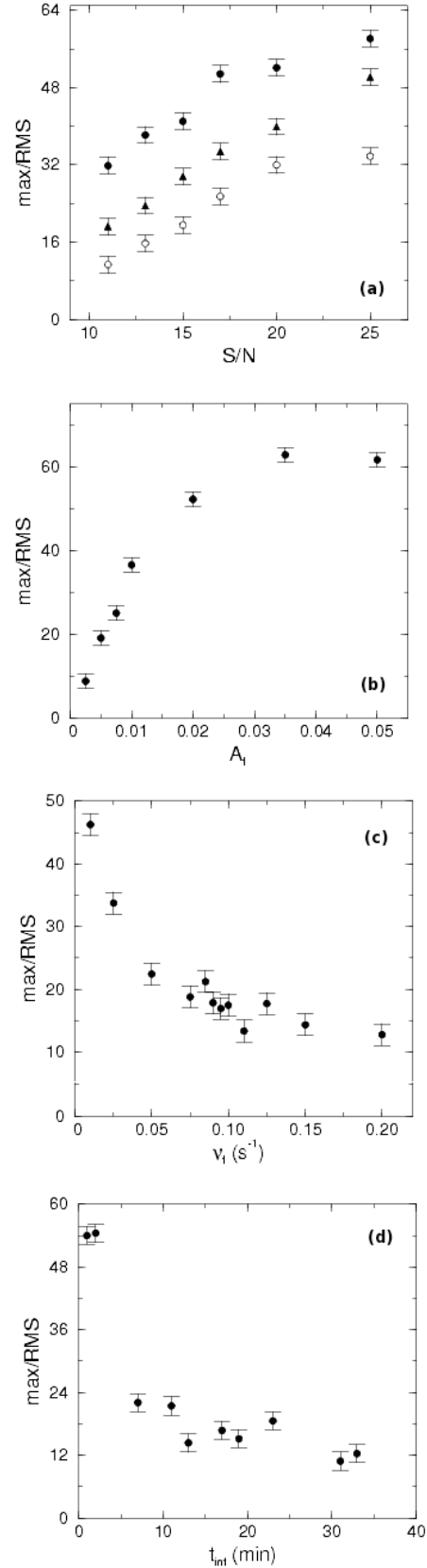}
 \end{center}
 \caption{Behavior of the quality factor with (a) S/N ratio, (b) flickering amplitude, (c) flickering frequency and (d) integration time. In (a) the filled circles correspond to simulations with 1000 spectra and flickering amplitude 1\%, the triangles were found with 500 spectra and amplitude of 0.5\% of the total line flux and the empty circles correspond to simulations with 1000 spectra and amplitude of 0.5\%.}
 \label{fig2}
\end{figure}

The addopted quality criterion to quantify the detection of a flickering feature was the ratio between the maximum intensity in the flickering occurrence region and the RMS of a region without flickering, affected only by the noise. We addopted as a feature detection criterion a quality factor ($max/RMS$) higher than 10.

In the figures 2 and 3, the effect of each flickering parameter of our simulation is presented. The error bars for all the simulations were calculated from the standard deviation of the quality factors obtained from a series of 11 simulations.

From figure 2$a$ one can see that, higher the number of spectra used to calculate the flickering Doppler map, better is the structure detection capability. The flickering amplitude also affects the structure detection, as features with higher amplitude are detected more clearly. If we consider that features with $max/RMS > 10$ are detected, it can be noticed that, even using a high number of spectra to calculate the maps, the flickering with lowest amplitude could not be detected. On figure 3, the synthetic flickering maps of two particular points from figure 2 are shown. One can see that the feature detection is better for figure 3$b$ ($S/N=20$) than for figure 3$a$ ($S/N=11$). The flickering amplitude effect on the Doppler maps is shown in figure 2$b$. It is clear that the detection is better for the case of higher flickering amplitudes. One also can see that low amplitude flickering information is lost more easily than the high amplitude flickering one. The effect of the flare occurrence frequency is presented in figure 2$c$. We notice a decrease on the detection quality for high frequencies, where the combination of discrete flares tends to form a continuum emission instead of a series of discrete events. It is important to observe that, in the case where the frequency is small than 1 flare per integration time, the variance calculated in each phase box may not represent the variability properly. The integration time effect upon the synthetic flickering maps is shown in figure 2$d$. One can see from this graph that the detection is better for small integration times (with many flares on each integration time). For longer integration times, the flickering information is lost. We have not found any significative variation of the quality factor with the decay time-scale of each flickering flare.

On these simulations the flickering source can be also attributed to the reprocessing of radiation at the illuminated face of the secondary star. The application of this particular source model to the observations of V3885 Sgr was already made \citep{Rib07}. In this case, the secondary is considered spherical and the flickering source geometry is approximate. These assumptions are reasonable, due to the typical resolution of Doppler tomograms. The velocity field in this case is given by eq. 2 without the Keplerian velocity term.

\begin{figure*}
 \begin{center}
 \FigureFile(125mm,65mm){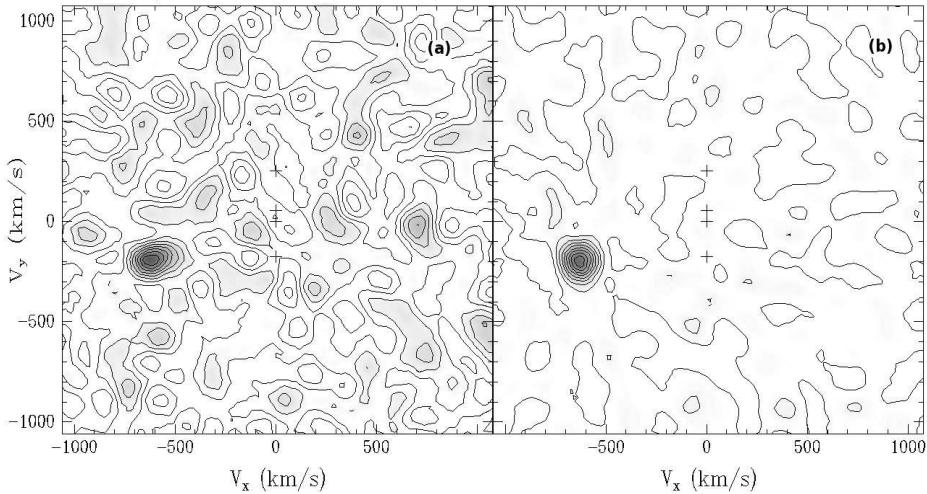}
 \end{center}
 \caption{Two synthetic flickering Doppler maps for 1000 spectra, flickering maximum amplitude of 0,5\% the total line flux and ($a$) $S/N = 11$ and ($b$) $S/N = 20$. Notice that the feature detection is much better on the second case. The crosses are, from top to bottom, the secondary's center of mass, the L1 Lagrange point, the system's center of mass and the primary's center of mass.}
 \label{fig3}
\end{figure*}

\subsection{Wind Simulations}

The wind is implemented in a 3d grid, below and above the disk. The wind emission process is chosen to be radiative recombination, with emissivity given by 
\begin{equation}
I_{wind}=\frac{1}{4\pi} h\nu\alpha_{eff}N_{e}^{2}
\end{equation} where $I$ is the wind emissivity, $\nu$ the frequency, $\alpha_{eff}$ the effective recombination coefficient and $N_{e}$ the electron density. For simplicity, the wind is considered isothermal. As only H$\alpha$ line profiles are simulated, other emission processes as scattering are not considered. If the temperature variation along the wind were considered, then a detailed temperature stratification on the disk must also be taken into account. The wind addopted geometry is biconic (figure 4), and its velocity law is given by the Castor \& Abbott velocity field \citep{Lon02} \begin{equation}
v_{poloidal}=v_{0}+\left(\beta\sqrt{\frac{2GM_{1}}{r}-v_{0}}\right)\frac{\left(l/R_{av}\right)^{\alpha}}{\left(l/R_{av}\right) ^{\alpha}+1}
\end{equation} where $\beta$ the wind terminal velocity scale factor, $M_1$ the primary mass, $v_0$ the initial wind velocity, $R_{av}$ the effective acceleration scale, $l$ is the distance of each wind element along the wind streamlines, $\alpha$ the acceleration coefficient and $r$ is the distance between a point in the wind and the white dwarf. The ionization was considered to be partial for H and He and up to the CNO elements. The disk angular momentum conservation through the wind is also considered by \begin{equation}
v_{keplerian}(r)=v_{keplerian}(r_{d}) \frac{r_{d}}{r sin(\theta_{v})} 
\end{equation} where $r_d$ is the radius of each wind element projected over the disk and $\theta_v$ is the elevation angular coordinate.

As the single peaked line profiles often seen in some CVs could not be generated only by varying the wind parameters, an optically thick wind was implemented. The line self-absorption through the wind was parameterized by the line optical depth $\tau$, which relates to the emitted radiation by
\begin{equation}
 I_\nu(s)=I_\nu(s_0) e^{-\tau}
\end{equation}
where $I_\nu(s_0)$ is original the wind emission and $I(s)$ the attenuated wind emission. Scattering effects were not considered.

The velocity of each wind element projected on the line of sight is calculated. If the element is not occulted by the optically thick disk, its emissivity is added to the disk spectra. If the wind has $\tau \neq 0$ the absorption along the wind region of the disk emission is also considered.

The result of these simulations are line profiles of Keplerian accretion disks with wind inside the primary's Roche lobe. The simulation can be repeated at many orbital phases, allowing us to construct Doppler maps with these synthetic spectra.

In this work the line profiles were simulated considering only the wind, i.e, the disk emission is not included, varying only the optical thickness to verify the effect of this parameter over the line profiles (figure 5). From figure 5 one can see that, as optical thickness is increased the line profile changes from double-peaked to single-peaked. In this simulation, we find that the line profile becomes single-peaked for $\tau$ values higher than 10. The effect of optically an thick wind on the Doppler maps is presented on figure 6. The optically thin Doppler map (figure 6$a$) is not filled at low velocities, while the optically thick wind Doppler map (figure 6$b$) presents a significant low velocity emission.

\begin{figure}
 \begin{center}
 \FigureFile(65mm,65mm){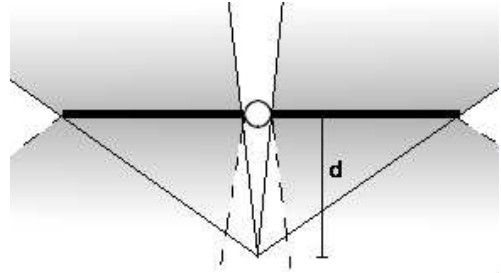}
 \end{center}
 \caption{Illustration of the wind biconic geometry. The white dwarf is the open circle at the center, the disk is the black thick line. The impact parameter $d$ is the distance between the white dwarf and the wind focus at each side of the disk. The wind cone is limited by the inner and outer disk rim, and it is presented as the gray gradient region. The wind cone at the other disk side is limited by the dashed line.}\label{fig4}
\end{figure}

\begin{figure}
 \begin{center}
 \FigureFile(85mm,85mm){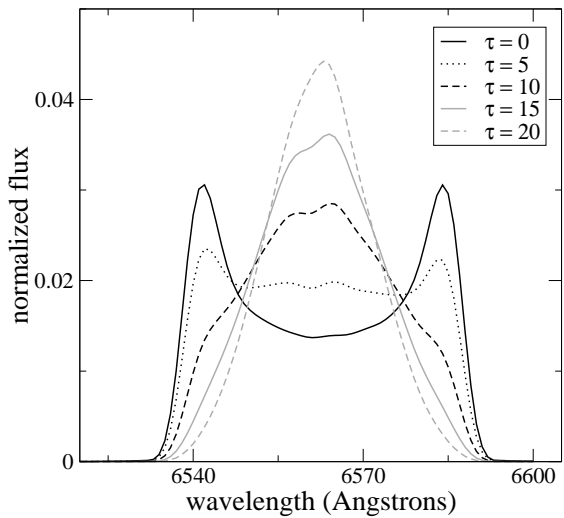}
 \end{center}
 \caption{Effect of optical depth on the wind line profiles. For optical depths higher than approximately 10 the lines are single peaked.}
 \label{fig5}
\end{figure}

\begin{figure}
 \begin{center}
 \FigureFile(75mm,75mm){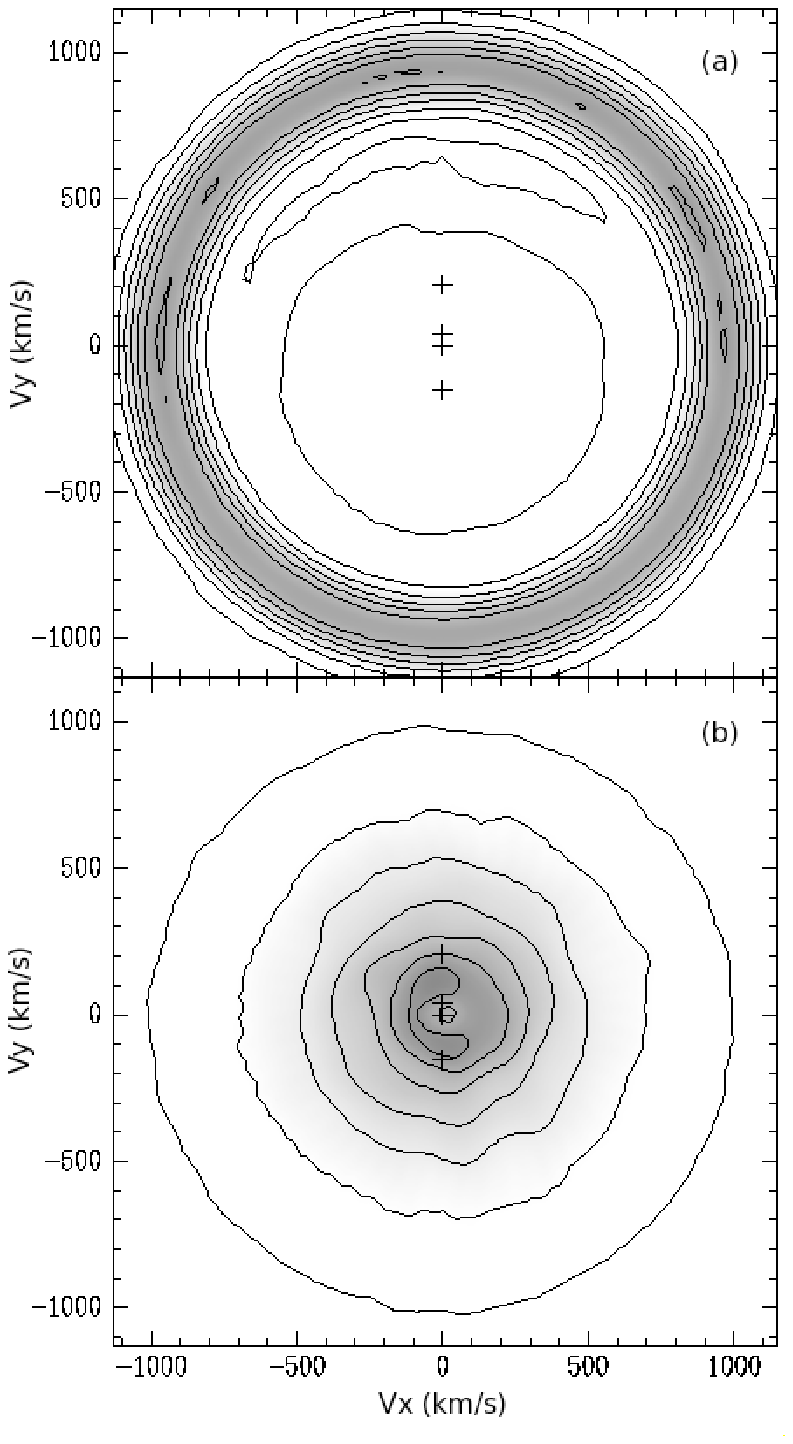}
 \end{center}
 \caption{Doppler maps for accretion disk simulations including ($a$) an optically thin wind and ($b$) an optically thick wind with $\tau_{line}$ = 20 (see text).}
 \label{fig6}
\end{figure}

The effect of the other wind simulation parameters (see eq. 6) on the optically thick wind line profiles are presented on figure 7. These simulations were held with a fixed line optical depth $\tau = 12.5$. The impact parameter $d_v$ is what determine the collimation of the wind. Higher the $d_v$ value is, more collimated is the wind. From figure 7$a$ one can see that a more collimated wind produces a smaller flux and a narrower profile. The acceleration coefficient $\alpha$ controls the wind acceleration. In figure 7$b$ one can see that a more accelerated wind generates most intense line profiles. One interesting point about this parameter is that, for low values of $\alpha$, a deficit in the emission at the red side of the line profile is perceived. This can be understood as the wind region opposite to the observer (redshifted mainly by the poloidal velocity component) having its radiation more attenuated by the absorption than the side in the direction of the observer (blueshifted). If only the Keplerian wind velocity is considered, the attenuation would be equal in both red and blue sides. The effect of terminal velocity scale factor $\beta$ on the line profiles (figure 7$c$) is that a higher value of $\beta$ yields line profiles with higher flux. A higher line flux is also the result of an increase of the effective acceleration scale $R_{av}$ (figure 7$d$) and a decrease on the initial wind velocity $v_0$ (figure 7$e$). In these last three cases the line profile shape does not present a significant change with the variation of each parameter, only the line flux does.

\begin{figure}
 \begin{center}
 \FigureFile(80mm,80mm){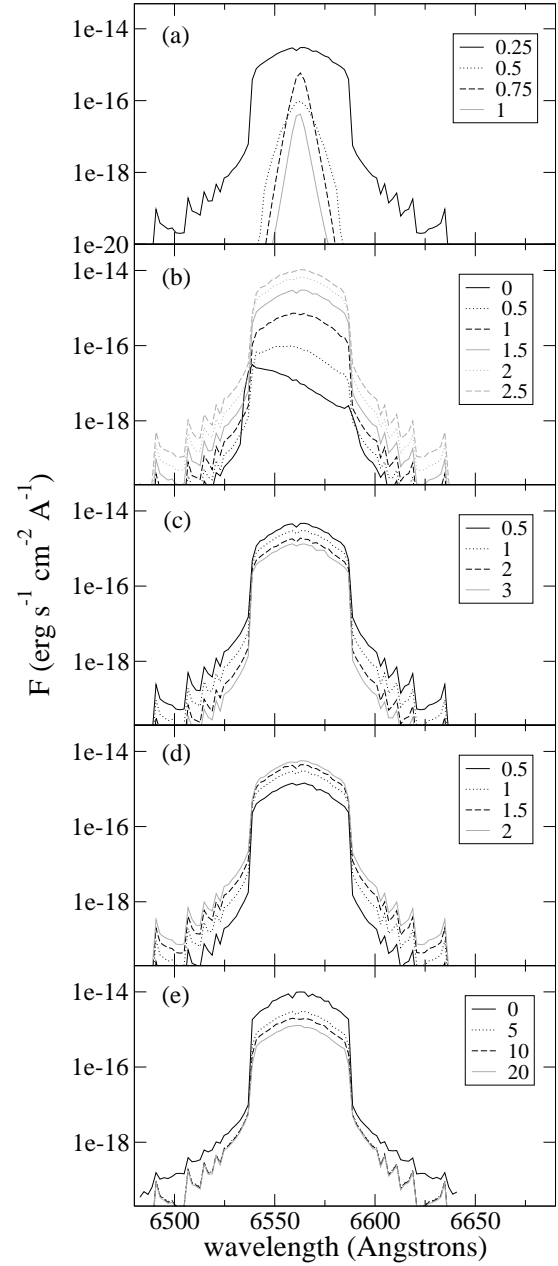}
 \end{center}
 \caption{Effect of the wind parameters on the wind line profiles, for a line optical depth $\tau = 12.5$. The parameters are ($a$) impact parameter, ($b$) acceleration coefficient, ($c$) terminal velocity scale factor, ($d$) effective acceleration scale and ($e$) initial wind velocity. The high frequency features that appear mainly in the line wings are effect of the finite grid addopted in the simulations.}
 \label{fig7}
\end{figure}

Another physical scenario that produces single-peaked line profiles and consequently low velocity emission on Doppler maps is the emission by stationary material inside the Roche lobe. In this case, the only velocity component is orbital and the density of the material inside the Roche lobe is assumed to be constant. The line profiles generated by emission of stationary material inside the primary's Roche lobe are similar to the ones produced by high optical depth winds. In the present simulations wind and/or emission by stationary material can be considered.

\begin{figure}
 \begin{center}
 \FigureFile(75mm,75mm){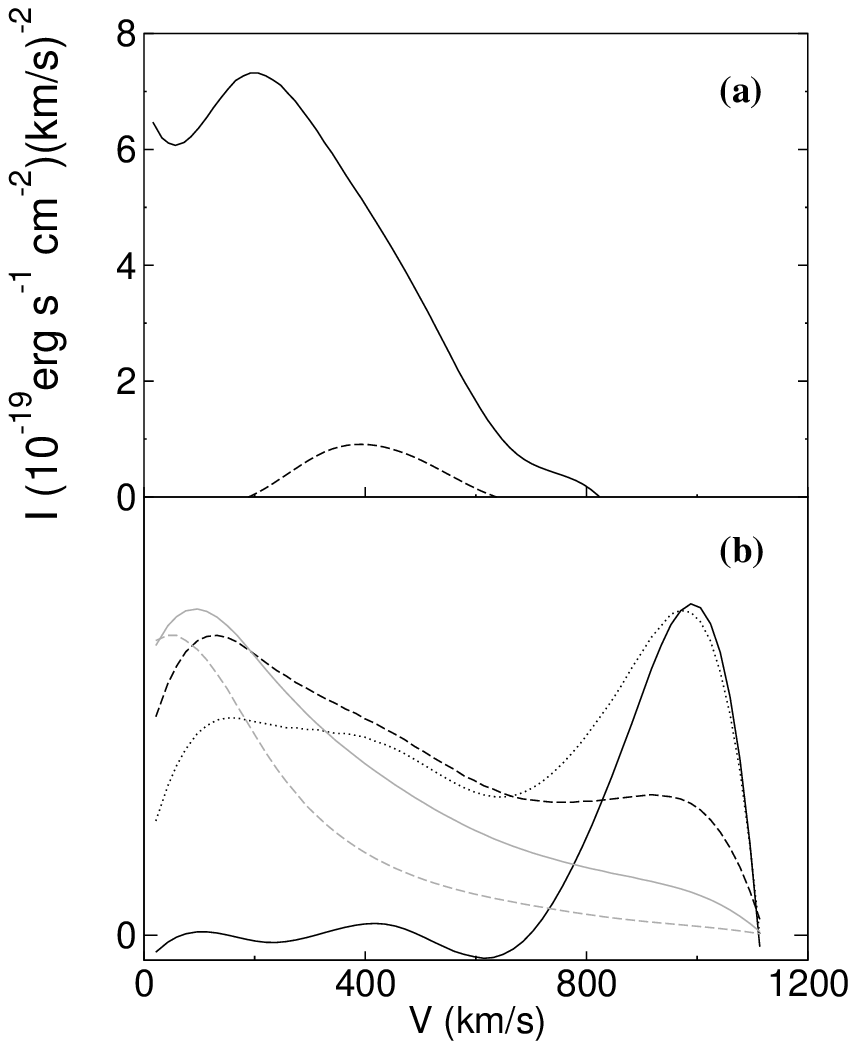}
 \end{center}
 \caption{Velocity profiles from ($a$) H$\alpha$ (\textit{solid curve}) and HeI $\lambda$6678 (\textit{dashed curve}) V3885 Sgr Doppler maps and from ($b$) simulated wind Doppler maps. On ($b$) the profiles are presented for some line optical depth values: $\tau=0$ (\textit{solid black curve}), $\tau=10$ (\textit{dotted black curve}), $\tau=15$ (\textit{dashed black curve}), $\tau=20$ (\textit{solid gray curve}) and $\tau=25$ (\textit{dashed gray curve}). The V3885 Sgr H$\alpha$ line velocity profile has a shape similar to those simulated with $15 \lesssim \tau \lesssim 20$.}
 \label{fig8}
\end{figure}

A test application was performed to verify the plausibility of the wind hypothesis to produce low velocity emission on observed Doppler maps. Wind Doppler maps were simulated with some different values of line optical depth. From each map a velocity profile was obtained by calculating the mode along rings centered on the primary. These velocity profiles were compared to the V3885 Sgr Doppler map velocity profile \citep{Rib07} (figure 8). From figure 8 one can see that the V3885 Sgr H$\alpha$ velocity profile has a shape similar to the simulated wind map velocity profile for high line optical depth %($20 \lesssim \tau \lesssim 25$). 
The velocity of the peak does not coincide because it depends on the dynamical parameters of the wind.

\begin{figure}
 \begin{center}
 \FigureFile(80mm,80mm){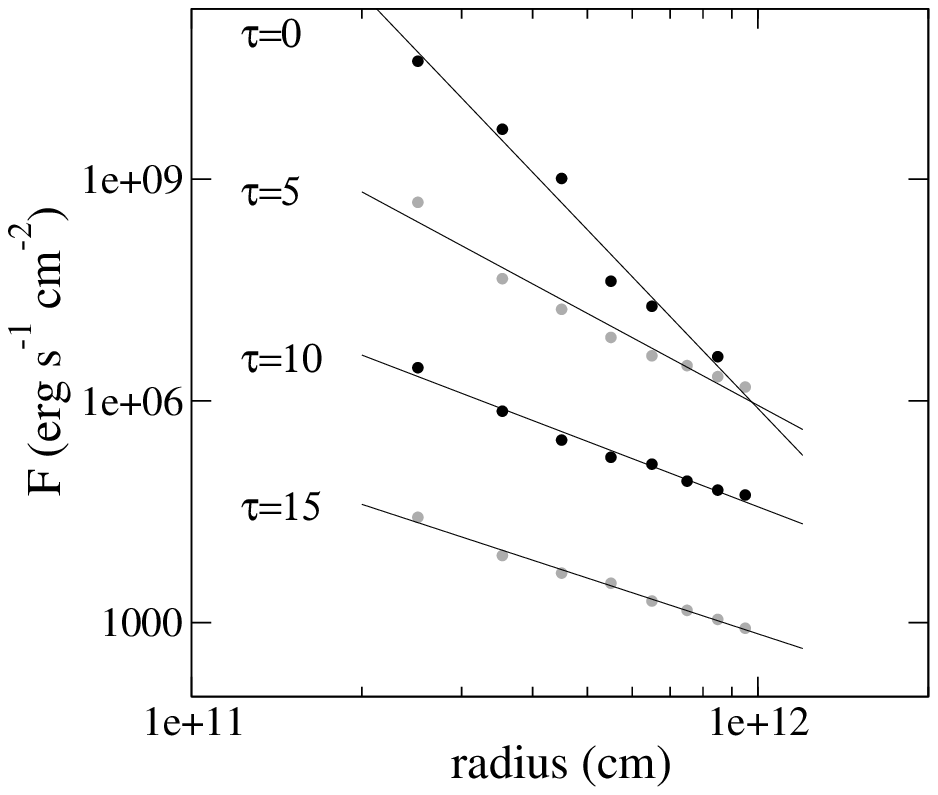}
 \end{center}
 \caption{Radial emissivity profiles from simulated H$\alpha$ wind Doppler maps. The data are displayed with alternated colors for each $\tau$ value for clarity}
 \label{fig9}
\end{figure}

We also analyzed the effect of the line optical depth on the radial emissivity profiles. We generated a series of simulated wind line profiles calculated with some fixed values for the line optical depth. Using those series Doppler maps were computed for each case. The disk contribution to the total emissivity was not included. The Doppler maps were transformed to the position space assuming a Keplerian velocity regime to obtain the radial emissivity profiles. The modal emissivity value was then calculated along several rings centered on the primary. Finally, a power law was fitted to these points (figure 9). The indexes of these radial power laws as a function of the line optical depth are given on table 1.

\begin{table}[htbp]
\begin{center}
\setlength{\belowcaptionskip}{10pt}% espaço entre caption e tabela
\caption{\it Power law indexes of the radial emissivity wind profiles as a function of the line optical depth}
\begin{tabular}{cc}
\hline
$\tau_{line}$ & power law index\\
\hline
0 	& -8,0 \\
5 	& -4,1 \\
10 	& -2,9 \\
15	& -2,5 \\
20	& -2,0 \\
25	& -1,6 \\
\hline
\end{tabular}
\end{center}
\end{table}

From table 1 we can conclude that the radial emissivity gets less centrally concentrated as we have higher line optical depths in the wind. From figure 9 one can see that, as expected, the flux is smaller for higher optical depths, but it is rather low for the $\tau_{line}$=15 case. This is expected as the emission measure gets too small in the high optical depth regime. The flux is also dependent on the other parameters considered on the simulations, as the electronic density or the dynamical parameters of the wind. The next step of our work is analyze the behavior of the simulated velocity and radial emissivity profiles with these parameters and compare the real data profiles with the simulated ones to constrain these dynamical parameters using a multiparametric optimization.

\section{Discussion}

In all flickering tomography studies the integration time must be small enough to preserve the flickering information. On the other hand, it must be long enough to ensure a good S/N ratio. One limitation of the Doppler tomography technique is that the flickering from the inner parts of the accretion disk could not be mapped with this method, as the information from a proportionally small accretion disk region is spread over a large region on the Doppler map. In addition, most of the emission from the inner parts of the accretion disk falls on the line wings, the profile region that is most affected by noise and continuum subtraction errors.

One possible extension of the flickering simulation study is to compare the observed flickering maps and simulated ones aiming to constrain the flare energy. The difficult of this study is that the flickering simulation has many parameters and that the noise is superposed to the intrinsic stochastic variation on the observed data. The flickering in the present work is simulated considering only one occurrence frequency, more frequencies can be included in future studies.

When the flickering from the secondary star is simulated, it is supposed to be due to the flickering from the disk being reprocessed on the illuminated face of the companion. To produce a more fiducial simulation, the effect of the illumination itself must be considered on the code. The effects of the illumination over the disk line profile is not included in the present version of the code.

On both the wind and flickering simulations, the disk emission is given by a power law emissivity radial profile. The present simulation code can be combined with an disk model code to improve the description of the disk emission properties. One obstacle to do this is that longer computational times will be necessary to simulate the spectra used on the construction of the synthetic Doppler maps. On the wind simulations, the wind was considered isothermal. In the case of a disk model combined with the code, the temperature stratification of the wind can be otherwise included.

The SW Sex stars are cataclysmic variables with peculiar characteristics. Among others, they present single peaked line profiles where a double peaked line profile is expected and their Doppler maps are dominated by emission at low velocity. This is occasionally observed even on systems which are not classified as SW Sex CVs. Many scenarios have been proposed to explain this behavior: accretion disk winds \citep{Hor86}, magnetic accretion and mass loss via a propeller mechanism \citep{Hor99}, a gas stream that does not impact the disk completely at the hot spot region and flyby over the disk \citep{Hel94}, accretion on the magnetic poles of a synchronous primary \citep{Cas96} and shock along the stream trajectory over the disk \citep{Gro00}. The present simulations and their comparison with observed Doppler maps may help to identify the origin of SW Sex phenomenon. In particular, the reconstructions of a disk with an optically thick wind suggest that disk mass loss should be at least part of the solution to the SW Sex problem.

\section{Conclusions}

A method for simulating accretion disk Doppler maps including wind and flickering is presented. The detection quality of the flickering raising from a small region of the disk was analyzed. High S/N and high time resolution spectra are needed to obtain information from flickering tomograms. The information from low amplitude and high frequency flickering is lost first. The flickering information is also lost if long integration times are used. The flickering from the inner part of the accretion disk could not be mapped with the Doppler tomography technique.

Wind simulations were also held, and the effect of each simulation parameter over the wind line profiles is presented. It was necessary to consider an optically thick wind, with optical depth higher than 10 to generate single peaked H$\alpha$ line profiles. Diffuse emission inside the primary Roche lobe is other mechanism that generate these single peaked profiles and centrally concentrated Doppler maps.

\section*{Acknowledgments}

This work is based on data obtained at LNA/CNPq and Cerro Tololo observatories. F.M.A.R is grateful from support from FAPESP fellowship 01/07078-8 and 06/03308-2. MD acknowledges the support by CNPq under grant \#304043.

%\section*{References}

\end{document}